\documentclass{aa} 
\usepackage{natbib}
\usepackage{graphicx} 
\usepackage{txfonts}
\usepackage{longtable} 
\usepackage{supertabular} 
\begin{document}

\newcommand{\rsc}{$r_{cs}$} 
\newcommand{\ntot}{$N_{tot}$}
\newcommand{\nbkg}{$N_{bkg}$} 
\newcommand{\nmem}{$N_{mem}$}
\newcommand{\nfield}{$N_{field}$}
\newcommand{\parchi}{$\chi^2_S$} 
\newcommand{\parchithresh}{$\chi^2_{S,threshold}$}
\newcommand{\be}{\begin{equation}} 
\newcommand{\ee}{\end{equation}}

\title{U-band photometry of 17 WINGS clusters}
\author{A. Omizzolo\inst{1,2} \and G. Fasano\inst{2} \and D. Reverte Paya\inst{3} \and C. De Santis\inst{4,5} \and A. Grado\inst{6} \and D. Bettoni\inst{2} \and B. Poggianti\inst{2} \and M.D'Onofrio\inst{7} \and A. Moretti\inst{7} \and J. Varela\inst{8} \and J. Fritz\inst{9} \and M. Gullieuszik\inst{2} \and A. Cava\inst{10} \and A. Grazian\inst{11} \and M. Moles\inst{8}}

\offprints{alessandro.omizzolo@oapd.inaf.it}

\institute{Specola Vaticana, 00120 Vatican City State \and 
INAF- Padova Astronomical Observatory, Vicolo Osservatorio 5, 35122 Padova, Italy \and GRANTECAN S. A., Centro de Astrof\'isica de La Palma, C/Cuesta de San Jos\'e s/n, 38712 Bre\~na Baja (La Palma), Spain \and INFN, Sezione di Roma Tor Vergata, I-00133 Rome, Italy \and Department of Physics, University of Rome Tor Vergata, I-00133 Rome, Italy \and INAF- Napoli Astronomical Observatory, Salita Molariello 16, 80131 Napoli, Italy \and Department of Physics and Astronomy, Vicolo Osservatorio 2, 35122 Padova, Italy \and Centro de Estudios de F\'isica del Cosmos de Arag\'on (CEFCA), Plaza San Juan 1, planta 2, E-44001 Teruel, Spain \and Sterrekundig Observatorium Vakgroep Fysica en Sterrekunde Universiteit Gent, Krijgslaan 281, S9 9000 Gent \and Observatoire de Gen{\`e}ve, Universit{\'e} de Gen{\`e}ve, 51 Ch. des Maillettes, 1290 Versoix, Switzerland \and INAF- Roma Astronomical Observatory, I-00040 Monteporzio Italy }

\date{Received / Accepted}

\abstract
{This paper belongs to a series presenting the \emph{WIde Field
    Nearby Galaxy-cluster Survey} (WINGS). The WINGS project has
  collected wide-field, optical (B,V) and near-infrared (J,K) imaging,
  as well as medium resolution spectroscopy of galaxies in a sample of
  76 X-ray selected nearby clusters (0.04 $<z <$ 0.07), with the aim
  of establishing a reference sample for evolutionary studies of
  galaxies and galaxy clusters.}
{In this paper we present the U-band photometry of galaxies and stars
  in the fields of 17 clusters of the WINGS sample. We also extend to
  a larger field of view the original B- and V-band photometry (WINGS-OPT)
  for 9 and 6 WINGS clusters, respectively.}
{We use both the new and the already existing B-band photometry to get
  reliable (U-B) colors of galaxies within three fixed apertures in
  kpc. To this aim, in the reduction procedure we put particular care
  in the astrometric precision. Since not all the observations have
  been taken in good transparency conditions, the photometric
  calibration was partly obtained relying on
  the SDSS and WINGS-OPT photometry for the U- and optical bands,
  respectively.}
{We provide U-band (also B- and V-band, where possible) total
  magnitudes of stars and galaxies in the fields of clusters. Just for
  galaxies, the catalogs also provide geometrical parameters and
  carefully centered aperture magnitudes. The internal consistency of
  magnitudes has been checked for clusters imaged with different
  cameras, while the external photometric consistency has been
  obtained comparing with the WINGS-OPT and SDSS surveys.}
 {The photometric catalogs presented here add the U-band information
  to the WINGS database for extending the SED of the galaxies, in
  particular in the UV wavelengths which are fundamental for deriving
  the SFR properties.}


\titlerunning{WINGS: U-band photometry}
\authorrunning{A. Omizzolo et al.}
\maketitle

\section{Introduction}

In the currently standard cosmological paradigm, clusters accrete
individual galaxies and larger subclumps from their outskirts. In this
scenario, the infalling regions of clusters are naturally very
important, being the transition regions in which galaxies are subject
to a dramatic change of environment, feeling the effects of the high
density environment for the first time. A morphological transformation
of spirals into S0 galaxies appears to occur in clusters at z$>$0.2, most
likely driven by environmental effects (Dressler et al. 1997, Fasano
et al. 2000). The environment also appears to have a strong influence
on the star formation activity of disk galaxies in clusters at high
redshift, apparently suppressing it upon infall into rich clusters
(Balogh et al. 1997; Couch et al. 1998, 2001; Dressler et al. 1999;
Poggianti et al. 1999). Indeed, observations probing the star
formation, Hubble types and gas content of galaxies in clusters have
proved that the cluster outskirts are essential to understand galaxy
transformations
\citep{abraham,balogh,ellingson,solanes,pimbblet,treu,kodama1,kodama4,
  mcintosh}. In particular, several recent works have shown that in
the local Universe the correlation between star formation activity and
local density extends to very large clustercentric radii, well beyond
the cluster central regions \cite{lewis,gomez, balogh}. The assembly
of clusters, by itself, seems to be able to suppress the star
formation, as suggested by the detection of post-starburst galaxies at
the interface of cluster infalling substructures \cite{poggianti2}.

In the last years many high quality (HST) observations have been
devoted to the study of clusters at intermediate and high redshift,
while for the local volume, till a few years ago, Virgo, Fornax and
Coma clusters constituted the main reference sample. To fill in this
gap, we started the WIde Field Nearby Galaxy-cluster Survey
\citep[WINGS][hereafter Paper-I]{fasano2}. This survey has
focused on clusters located in the redshift range 0.04 -- 0.07 and has
collected wide-field optical (B,V) imaging \citep[][WINGS-OPT]{varela}
for a sample of 76 clusters selected from three X-ray flux limited
samples compiled from ROSAT All-Sky Survey \citep{ebeling1,ebeling2,ebeling3}. In addition,
multi-fiber, medium resolution spectroscopy and near-infrared (J,K),
wide-field imaging have been obtained for 48 and 28 WINGS clusters,
respectively \cite{cava,tiziano}.

To complement the WINGS database with U-band imaging, we have gathered
observations for a subsample of 17 WINGS clusters using three
different telescopes equipped with different wide-field cameras and a
few archival data.

These observations will allow to study in detail
the star formation activity in a statistically significant sample of
cluster galaxies. As the integrated spectrum of a galaxy is more and
more dominated by young stars going to shorter wavelengths, U-band
data is by far more sensitive to any current or recent star formation
than any of the other broad-bands available \cite{kennicutt, barbaro}. Our
U-band estimates of the current star formation are truly
integrated values (though dust-affected), to be compared with the estimates based on our
optical spectroscopy, which samples only the very central
regions of each galaxy ($1.6\arcsec$/2.6$\arcsec$). Moreover, the
spatial distribution of the U-band emission within galaxies
greatly helps to discriminate between the various physical
processes, by revealing if star formation preferentially is suppressed
and/or enhanced in the central and outskirt regions of galaxies. In
particular, the distribution of the star formation activity (SFA) in
galaxies residing in the cluster outskirts (infalling region) could
reveal if they are affected by ram pressure stripping (SFA in the
outer regions), by shock-induced star formation related to the galaxy
infalling into the cluster (SFA on the impact edge of the galaxies),
or by centrally driven starbursts due to tidal repeated encounters in
the cluster potential (SFA in the nuclear regions). Moreover, these
observations should allow us to establish how the SFA correlates with
galaxy morphology (from the V-band imaging), mass (from K-band data) and spectra, as
well as with the environment (local density and cluster properties).

The paper is organized as follows. Section 2 describes our
observations and data reduction. In Section 3 we present the catalogs
and in Section 4 the data quality is analized. Through the paper we
use the following cosmology: $H_0$ = 70 km $s^{-1} Mpc^{-1}$, $\Omega
_m$ = 0.3 and $\Omega\ _\Lambda$ = 0.7.  The magnitudes are in the Vega system.

\section{Observations and data reduction}

The data presented in this paper are based on observations obtained
with three different wide-field cameras (see Table~1): {\it (i)} the
90prime camera at the 90 inches BOK telescope (90prime@BOK, Kitt
Peak); {\it (ii)} the Wide Field Camera at the 2.5m Isaac Newton
Telescope (WFC@INT); {\it (iii)} the Large Binocular Camera at the
Large Binocular Telescope (LBC@LBT). For one cluster (Abell~970) we
used imaging data from the WFI@MPG (ESO2.2 archive). All clusters have
been imaged in the U-band. Many clusters have also been imaged in the
optical (B,V) bands.

\begin{table}
\begin{center}
\begin{tabular}{cccccc}
\hline
Telescope (Camera) & Pixel scale  & e$^{-}$/ADU & RON e$^{-}$ & FOV\\
\hline
INT (WFC) & 0.333" & 2.8 & 6.2 & 34'x34'\\
BOK (90prime) & 0.450" & 1.71 & 12 & 90'x90'\\
LBT (LBC) & 0.226" & 2.022 & 11.45 & 23.6'x25.3'\\
MPG (WFI) & 0.238" & 2.0 & 4.5 & 34'x33' \\
\hline
\end{tabular}
\caption{The cameras.}
\end{center}
\end{table}

\begin{table}
\begin{center}
\label{log}
\begin{small}
\begin{tabular}{cccccc}
\hline
Date  & Telescope & N\\
\hline
February 27-29 2000 & WFI@MPG & 1\\
May 10-14 2005 & WFC@INT & 8\\
June 20-22 2006 & 90prime@BOK & 2 \\
November 22-23 2006 & 90prime@BOK & 4 \\
March 12 2007 & LBC@LBT & 1 \\
May 19 2007 & LBC@LBT & 1 \\
June 04 2008 & LBC@LBT & 2 \\
June 07 2008 & LBC@LBT & 1 \\
June 08 2008 & LBC@LBT & 2 \\
\hline
\end{tabular}
\end{small}
\caption{The runs. For each run: (1) the date; (2) the name of
the telescope;  (3) number N of clusters observed.}
\end{center}
\end{table}

In Table~\ref{log} we report the observing log, one row per run (per
night in the case of the LBC observations), each row including the
number of imaged clusters.  In Table~3 the list of the observed
clusters is reported. Besides the average coordinates and redshifts of
the clusters, in column 5 we list the telescopes used to image each
cluster. The BOK observations always include the U-, B- and V-bands,
while the INT and MPG observations just include the U-band. For the
clusters imaged with LBT, we list in parenthesis the filters used
for the imaging. The {\it FWHMs} for each cluster in each filter are
given in the headers of the catalogs (see in Figure~\ref{catalog}
the first group of rows of the header). With the WFC, 90prime and LBC
cameras we have imaged 8, 6 and 6 clusters, respectively. Some
clusters have been observed with two or three instruments for a total
of 15 clusters observed in the U band. To these observations we added
U-band imaging of the cluster Abell 970, taken from the WFI@MPG
(ESO2.2 archive). The data sets coming from the four instruments (WFC,
90prime, LBC and WFI) span a time interval of about four years and
reflect different proposal strategies (observing constraints and
requirements), also depending on the available observing
time. Therefore, because of the quite heterogeneous instrument sets
and weather conditions, we used different reduction strategies for the
different cameras, as outlined in the following sub-sections.

\begin{table}
\begin{center}
\begin{tabular}{ccccc}
\hline
ID  & $\alpha{J2000}$& $\delta{J2000}$ & z & Telescope \\
 \hline
A0119  & 00 56 21 & -01 15 & 0.0444 & BOK  \\
A0970  & 10 17 34 & -10 42 & 0.0580 & MPG  \\
A1291  & 11 32 21 & 55 58 & 0.0527 & INT \\
A1668  &  13 03 46 & 19 16 & 0.0634 & LBT (U)\\
A1795  & 13 48 52 & 26 35 & 0.0622  & BOK, INT, LBT (U,B) \\
A1831 & 13 59 15 & 27 58 & 0.0612   & INT \\
A1983 & 14 52 59 & 16 42 & 0.0444   & INT \\
A1991 & 14 54 31 & 18 38 & 0.0586   & LBT (U,B) \\
A2107 & 15 39 39 & 21 46 & 0.0411   & BOK, LBT (U,B) \\
A2124 & 15 44 59 & 36 06 & 0.0654   &INT, LBT (U,B) \\
A2149 & 16 01 35 & 53 55 & 0.0675  & BOK \\
A2169 & 16 14 09 & 49 09 & 0.0579   & INT \\
A2399 & 21 57 13 & -07 50 & 0.0582  & BOK \\
RXJ1022 & 10 22 10 & 38 31 & 0.0534 & BOK \\
RXJ1740 & 17 40 31 & 35 39 & 0.0430   & LBT(B) \\
ZwCl2844 & 10 02 36 & 32 42 & 0.0500 & INT \\
ZwCl8338 & 18 10 50 & 49 55 & 0.0473 & INT \\
\hline
\end{tabular}
\caption{The observed clusters.}
\end{center}
\end{table}

\subsection{INT observations}

We obtained WFC@INT imaging of 9 clusters during three useful nights
of the same observing run (May 10/12/13 2005). For the outline of the
camera we refer to Table~1 and Paper-I. The observations have been
taken, under generally good weather conditions, just in the U-band and
trying to match as much as possible the field of view of the B- and
V-band imaging already available in the WINGS-OPT survey. To cover the
gaps between the CCDs, at least three 20~min dithered exposures per
cluster have been done. During each night standard field exposures
have been secured to allow the photometric calibration.

Bias subtraction and flat-field corrections were separately performed
on each of the four CCDs of the WFC, while the mosaic-image of each
exposure was produced using the IRAF tool wfcmosaic. The mean error of
the astrometric solution, obtained for every exposure through the USNO
star catalog, turned out to be $\le$0.3 arcsec ($\sim$1~pixel) over
the field. The final image of each cluster was obtained by weighted mean
combination of the dithered exposures.

During the run, the system proved to be rather stable, while the
photometric calibration showed that the transparency did not change
significantly. The cluster A1991 was excluded from the final INT sample
since the outlined reduction procedure was not able to repair some
temporary failure of the acquisition system causing a lack of
homogeneity among the CCDs.

\subsection{BOK observations}

The 90prime camera mounted at the BOK telescope (Kitt Peak) during our
2006 observing runs was a mosaic of four CCDs separated in both
directions by very large inter-CCD gaps (about 15.76 mm or 1050
pixels). The edge-to-edge field of view, including the inter-CCD gaps,
was 1.16$\times$1.16 degrees, with a plate scale of 30.2"/mm or
0.45"/pixel.

In five nights, sharing two observing runs (see Table~\ref{log} for
details), we have imaged 8 clusters in the U-, B- and V-band. In
order to fill the large gaps between the CCDs, we shifted the
telescope in 5 different positions, thus covering the entire field of
view of the camera.

During each night, dome and twilight sky flats were obtained and
several photometric standard fields were imaged in each
photometric band and at different zenithal distances. Unfortunately,
in both observing runs the weather conditions were inclement. In
particular, the average seeing was about 2" and the sky transparency was
not good.

Due to the very large angular view provided by the 90prime camera, in
order to obtain a good enough astrometry over the whole field, we were
forced to adopt a more laborious procedure with respect to the INT
data reduction. In particular, relying on the stars of the USNO
catalog in a suitable (filter dependent) magnitude range, we first
obtained, for each filter and for each CCD, an average astrometric
solution relative to the geometrical center of the camera, staking
altogether the mosaic images obtained in that filter. Then, using 15
stars for each CCD (3 in every corner and 3 in the center), we
improved the astrometic solution of each CCD for each exposition and
translated it into the proper position on the sky. The $rms$ of the
angular distances between the BOK and USNO coordinates turned out to
be less than 0.25 arcseconds.

To obtain the final mosaic for every field, we used SWARP
\citep{bertin} by Terapix. Using our astrometric projection defined in
the WCS standard, for each cluster in each filter, SWARP resampled and
co-added the set of five dithered exposures, thus producing the final,
backgroud-subtracted image. Two of the eight imaged clusters (A2256
and RX0058) were observed under quite poor weather conditions. They 
have not been included in the final BOK sample.

Even though the standard field exposures were diligently processed as
we did for the scientific ones, they were just used to obtain the
relative photometric calibration within the 90prime field, i.e. to
estimate possible gain and linearity differences among the CCDs. In
fact, because of the poor transparency, we were forced in both runs
to perform the absolute photometric calibration (zero points and color
terms) relying on the WINGS-OPT catalogs \citep[][for the V- and
B-band]{varela} and on the Sloan Digital Sky Survey DR7 photometric
data \citep{aba}, suitably converted to the Johnson system
\citep[][for the U-band]{lupton}.

\subsection{LBT observations}

LBC fits images are Multi Extension Files (MEF) composed by a mosaic
of 4 CCDs of 4608x2048 pixels with a median plate scale of 0.225
''/pixels and a scientific field of view (FoV) of about 23.6x25.3
arcmin$^2$.  Therefore, in order to cover the field imaged by the
WINGS-OPT survey, five exposures per cluster were planned.

The observations of our clusters have been taken in service mode,
during the Science Demonstration Time (SDT) under variable
transparency and seeing conditions. The data set resulting from these
circumstances was not optimal, since in many cases the cluster field
coverage was incomplete (sometimes sparse), or the different exposures
of the same cluster were taken in different seeing and transparency
conditions.  At the end, just 6 clusters turned out to have enough
field coverage and seeing homogeneity to be included in the final LBT
sample. Three of them were imaged in both U- and B-band, while for two
clusters just the U-band imaging was available. Finally, for RX1740 we
just obtained B-band imaging.

The reduction has been performed by one of us (CD) using the standard
procedure devised by the LBC Team\footnote{\sl
  http://lbc.oa-roma.inaf.it/}.  In most CCD mosaic imagers,
electronic ghosts are present due to video channel's cross-talk and a
specifically designed software procedure (xtalk) has been used to
remove these features; for LBC the cross talk coefficient is about
3.0E-05.

The flat field images were derived from twilight sky data and a
calibration master flat was obtained by stacking a set of flat images
with a sigma clipping rejection algorithm with radial profile
normalized to unity in the center. The saturated pixels, chips' bad
signatures, cosmic ray events and satellite trails were masked using a
special derivative algorithms developed by the LBC Team.

The area of the LBC pixels is not constant over the entire FoV, due to
the effect of astrometric distortions (also called sky
concentration). To correct this feature and normalize the pixel area,
a sky-concentration image was applied as multiplicative factor. This
filter dependent correction image was produced by a specific software
that uses as input information the astrometric solution for the
specific filter. After this correction, it was possible to mask all
objects in the image and stack all frames within a fixed temporal
window (about 10 minutes), in order to allow a proper subtraction of
the small scale features of the background. This temporal window is
representative of the main sky background variation and features such
as fringes.  For the U- and B-band this procedure produced well
flattened images and no further processing was required.

The background subtracted images were corrected for object photometry
altered by sky-concentration effects dividing for the same correction
image used in the previous reduction step. In fact, the sky
concentration effects modify the surface brightness of the background,
producing a typical pin-cushion feature, but it does not alter the
integrated flux of extended objects. Therefore, each science image
must first be multiplied by the sky-concentration image to rectify the
background, and once it is properly subtracted, it must be divided by
the same sky-concentration image to correct the flux of stars and
galaxies to their original value. Being a prime focus camera, the
optical distortions of LBC are quite relevant. Still, after the above
outlined procedure, the quality of the PSF over the entire LBC FoV
turned out not to depend on the radial distance from the optical
center.

The astrometric solution was computed through a three-pass process by
the AstromC package \cite{radovich}: 1) the offset of the four chips
were computed by matching a catalog of objects found on the frame with
an astrometric catalog (usually USNOA B1.0 catalog); 2) an overall fit
was performed to obtain a local chip-to-chip astrometric solution
applying the calculated deformation map obtained with Stone (1997)
astrometric fields (and UCAC catalogs), as described in
Giallongo et al. (2008, G08 hereafter); 3) the final astrometric solution was computed on
the whole dithered image set thus providing a well minimized global
fit.

Since no photometric standard fields were imaged during the SDT runs,
the (provisional) photometric calibration was obtained using the
coefficients given in Table~2 of G08. These coefficients
are the result of several commissioning nights just devoted to the
photometric characterization of LBC@LBT. The nominal photometric
accuracy in the overall field is of the order of 0.01 mags. We refer
to G08 for further details about the LBC instrument and
the data reduction procedures. The final zero points have been checked
(and sometime improved) using the SDSS DR7 and the WINGS-OPT
databases for the U- and B-band, respectively.

\subsection{ESO WFI@MPG archival data}

Form the ESO archive we retrived U-band deep observations of one more
cluster, namely A970. The images were taken during the night of
Feb. 28, 2000.  They have been processed using the VST-Tube \cite{lino}
pipeline.  In particular, after bias and flat-field correction, to the
mosaic image a provisional absolute calibration was applied using a
few photometric standard field observed during the same night and
adopting the extinction coefficient given in the ESO La Silla WEB page
(0.48).

The same photometric standard fields were used to determine the
illumination correction map. This has been obtained by using a
generalized adaptive method (GAM) to interpolate the difference
between raw and standard magnitudes as a function of the position.
The GAM allows to obtain a well behaved surface also in case (like the
present one) the field of view is not uniformly sampled by the
standard stars. The illumination map image was then used to correct
the science images during the pre-reduction stage. After applying the
illumination correction the (raw - standard) magnitude difference is
reduced by a factor $\sim$2.

The gain harmonization among the CCDs and the astrometric solution
were obtained using SCAMP (Bertin 2007). The r.m.s. (along each axis)
of the pairwise differences between coordinates of overlapping
detections and between detection coordinates and coordinates of the
associated astrometric reference stars, turned out to be
$0.177\arcsec$ and $0.168\arcsec$, respectively.

\begin{table*}
\begin{tabular}{ccccccc}
\hline
Telescope (Camera) & FWHM  & Astrometry  & U$_{lim}$(90\%) & B$_{lim}$(90\%) & V$_{lim}$(90\%) \\
\hline
INT (WFC) & 1''.30 (0''.27) & $<$0''.3 & 22.3$\div$23.3 &  & \\
BOK (90prime) & 1''.93 (0''.23) & $<$0''25 & 21.3$\div$22.1  & 21.1$\div$22.6 & 20.0$\div$22.5 \\
LBT (LBC) & 1''.40 (0''.24) & $<$0''.22 & 22.6$\div$23.3 & 23.1$\div$23.6 & \\
MPG (WFI) & 0''.9 & $<$0''.19 & 22.1 &  &  \\
\hline
\end{tabular}
\caption{Quality of observations}
\end{table*}

Columns 2 and 3 of Table~4 list the median seeing ($rms$ in parenthesis) and the astrometric quality of the final mosaics for each camera.

\section{The Catalogs}\label{catalogs}

The source detection and extraction has been performed using
SExtractor \citep[][SEx hereafter]{bertin}.  For each mosaic frame,
with the proper seeing and photometric depth, we performed a number of
test runs of SEx to identify the most suitable values of the deblendig
parameters, trying to find compromise values running well for
different kinds of objects. In this way, we sacrificed the homogeneity
of the catalogs in order to obtain galaxy samples as complete as
possible. For each telescope (INT/BOK/LBT/ESO) we provide catalogs of
each cluster, including the magnitudes for all the available
photometric bands. SExtractor provided integrated (MAG\_AUTO)
magnitudes of sources, as well as the geometrical parameters of
galaxies and the automatic star/galaxy classifier. Where possible, the
geometrical and star/galaxy SEx parameters are referred to the
B-band. In case the B-band imaging is not available, they are referred
to the U-band imaging.

In the original, WINGS-OPT catalogs, particular care was devoted to
distinguish stars and galaxies \citep[see Section 2.3 in
][]{varela}. Therefore, we assumed that the star/galaxy classification
given therein is correct and, in each cluster, we adopt this
classification for the objects in common with the WINGS-OPT
catalogs. Moreover, the common objects have been used to compare the
continuous (from 0 to 1) star/galaxy SEx classifier of the new
catalogs with the binary star/galaxy classifier of the old WINGS-OPT
catalogs. In particular, using the star/galaxy, ellipticity and FWHM
parameters of the new SEx catalogs, for each camera of the present
survey, we have identified some empirical criteria to {\it tranfer}
the old classification to the objects not in common with WINGS-OPT. We
adopt this second-hand, indirect (binary) classification for the `new'
objects, which mostly reside in the outer cluster regions, not sampled
by the original WINGS-OPT survey. From the visual check of 200
detections (100 stars and 100 galaxies), randomly selected among the
`new' objects of all catalogs with magnitude $U_{Tot}<$22.5, it turned
out that this indirect, binary classification is correct for 83 and 75
stars and galaxies, respectively, while for 28 objects the visual
classification is uncertain. The percentages become higher (44/45 and
17/18 for stars and galaxies) if one considers just objects with
$U_{Tot}<$20.5.

In spite of the relatively high values chosen for the detection
threshold (1.5$\sigma_{bkg}$/pixel), the number of detected sources in
each filter turned out to be often quite large, due to the large
number of spurious detections. Therefore, where possible (i.e. for
clusters observed in two or three bands), we decided to retain in the
final catalogs just the sources detected in all the available
photometric bands, thus producing catalogs which include, for the
common objects, the total magnitudes SExtracted in the different
bands. 

As mentioned in the previous section, while for the INT and ESO
imaging the photometry can rely on their own absolute calibrations,
for the BOK and LBT imaging the absolute calibration has been obtained
relying on the SDSS and WINGS-OPT star magnitudes for the U-band and
for the optical (B,V) bands, respectively. When two or three bands
were available for a given cluster, this relative calibration also
takes into account the color terms derived from the provisional SEx
magnitudes. The comparison U-band magnitudes are obtained from the
$u,g,r$ SDSS (DR7) magnitudes, using the conversion formula by
Lupton (2005).

\begin{figure*}
\includegraphics[scale=0.24]{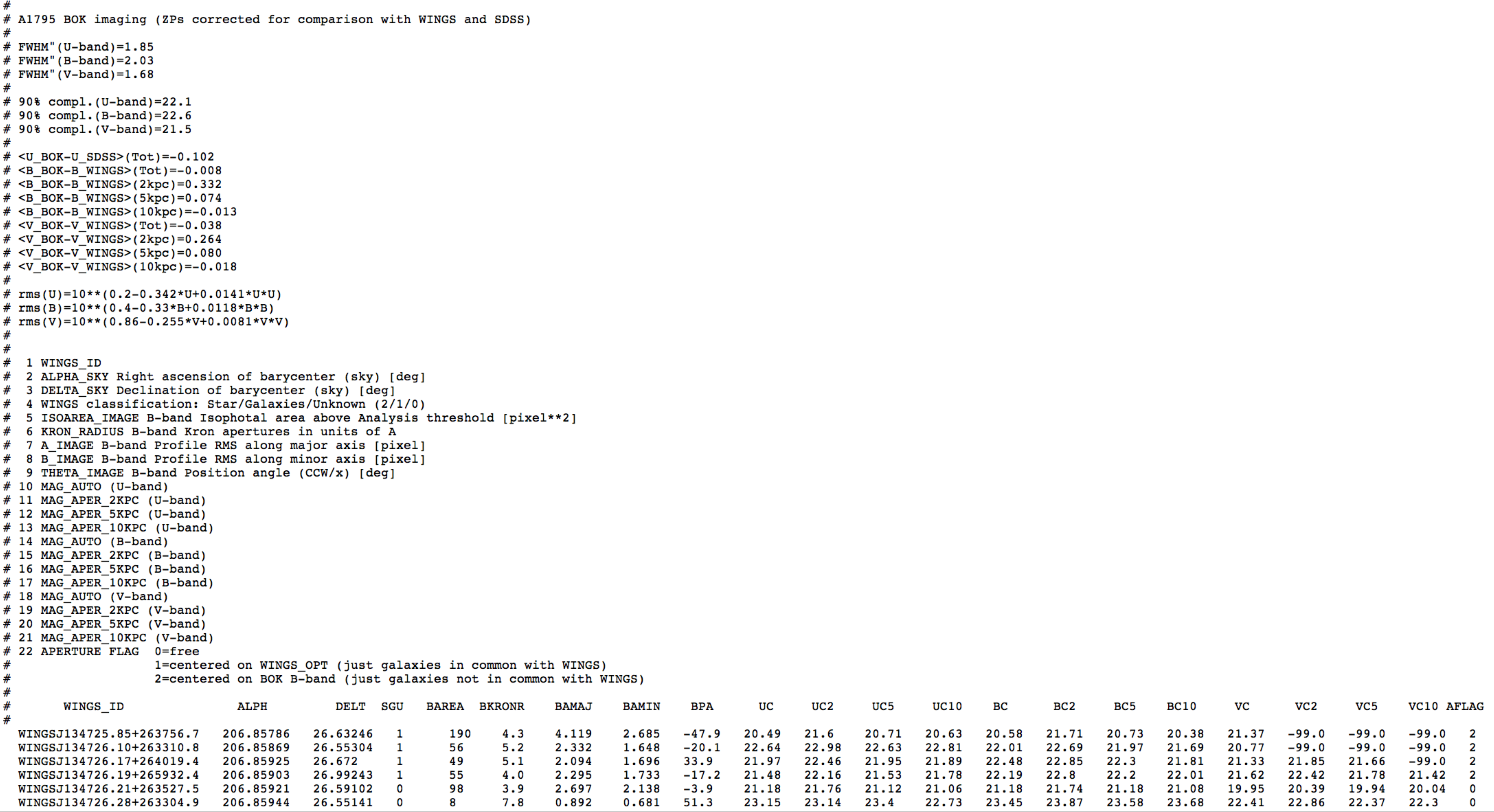}
\caption{Header and first rows of the catalog from BOK imaging of A1795}
\label{catalog}
\end{figure*}

Perhaps the most important target for the U-band photometry of large
galaxy samples in clusters is to determine their colors. It is well
known that, in order to obtain reliable color estimates, it is crucial
to measure the magnitudes inside apertures centered exactly at the
same points in the two bands. On the other hand, since galaxies
(especially of late-type morphology) may look differently in different
bands, the geometrical centers coming from automatic SExtraction of
sources turn out to be usually different in the different bands. To
overcome this problem, we have used a purposely devised script, based
on the IRAF apphot package, allowing us to measure the aperture
magnitudes exactly on the requested positions, with the requested
radii. Since in the WINGS-OPT catalogs we report the aperture
magnitudes within circular apertures of radii 2, 5 and 10~kpc, for
all galaxies in common with WINGS-OPT we measured the aperture
magnitudes in the U-band (and also in the optical bands, where
possible) adopting exactly the same centers and radii of the WINGS-OPT
survey. In the catalogs we report a special Aperture Flag (AFLAG, see
Figure~\ref{catalog}: last column), that for galaxies in common with
WINGS-OPT is set to 1. If the cluster has been imaged with some camera
in both the U- and B-band, we centered the U-band aperture magnitudes
of the galaxies not in common with WINGS-OPT on the apertures given by
SEx for the B-band imaging. For these galaxies AFLAG is set to 2 in
the catalogs. In the remaining cases (all stars and those galaxies not
in common with WINGS-OPT and just imaged in the U-band) AFLAG is set
to zero.

As an example of the catalogs presented in this paper,
Figure~\ref{catalog} shows the header and the first rows of the
catalog from the BOK imaging of A1795. The $rms$ of magnitudes as a
function of the magnitudes themselves have been estimated fitting
second order polinomials to the decimal logarithm of the magnitude
binned $rms$ of the differences between the total magnitudes of
galaxies and the comparison magnitudes from the WINGS and/or SDSS
surveys. These polinomials are reported in the headers of catalogs
(see Figure~\ref{catalog}: fourth group of rows in the header) and can
be used to assign individual errors to the magnitudes.

\begin{figure*}
\includegraphics[scale=0.7]{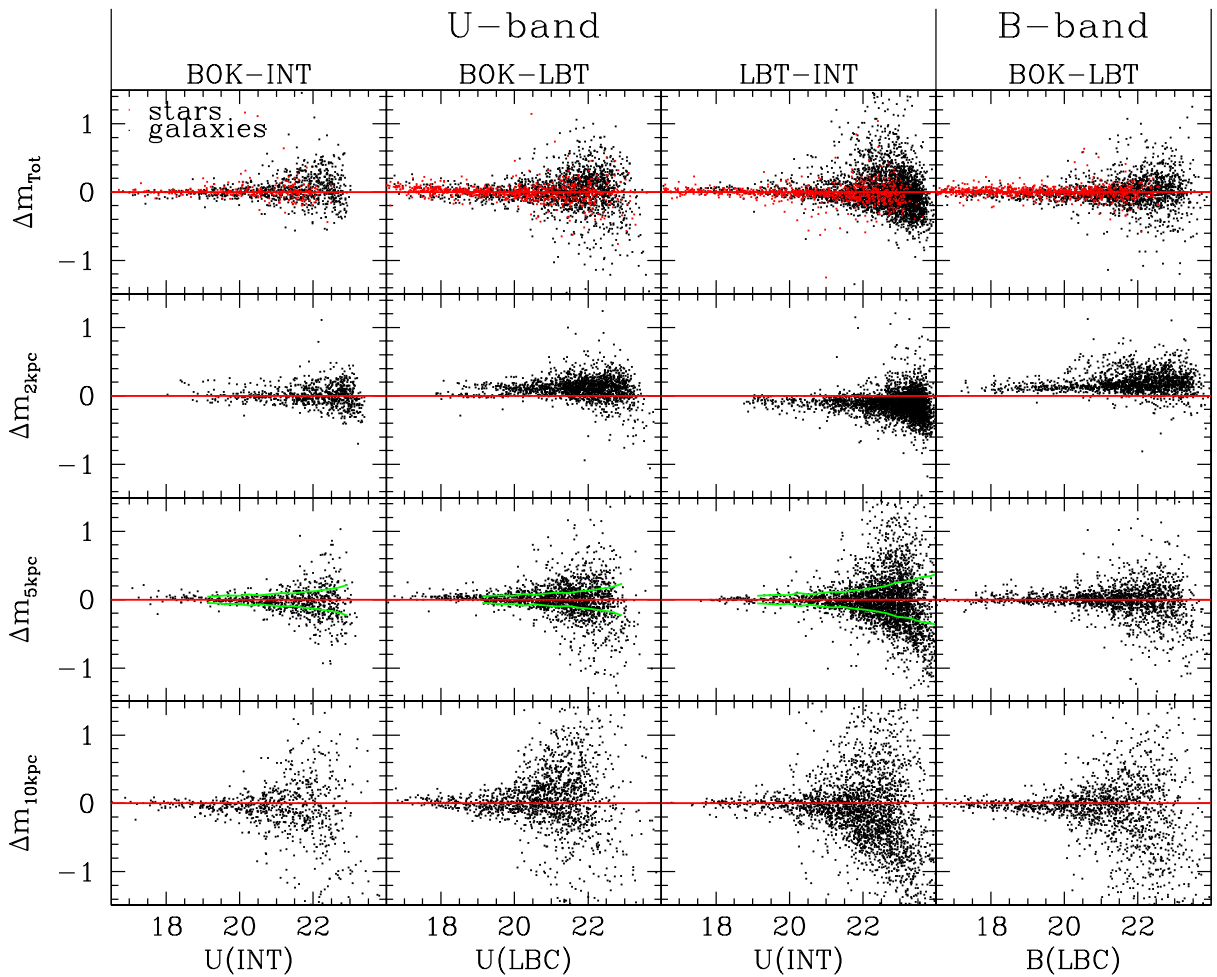}
\vspace{-2truecm}
\caption{Differences between magnitudes obtained using different
  cameras for common galaxies (black dots) as a function of the
  magnitude. The plots in the topmost row of the figure refer to the
  total magnitudes, while those in the remaining rows refer to the
  magnitudes within circular apertures of radii 2, 5 and 10~kpc, top
  to down. The plots in the rightmost column of the figure refer to
  the B-band, while those in the other columns refer to the U-band.
  The red dots in the plots of the topmost row of the figure (total
  magnitudes) refer to the stars. The green curves in the plots
  relative to the 5~kpc apertures (third row of the figure) illustrate
  the $rms$ expected according to the formulas reported in the headers
  of the catalogs (see text for details).}
\label{internal}
\end{figure*}
 
\begin{figure}
\includegraphics[scale=0.45]{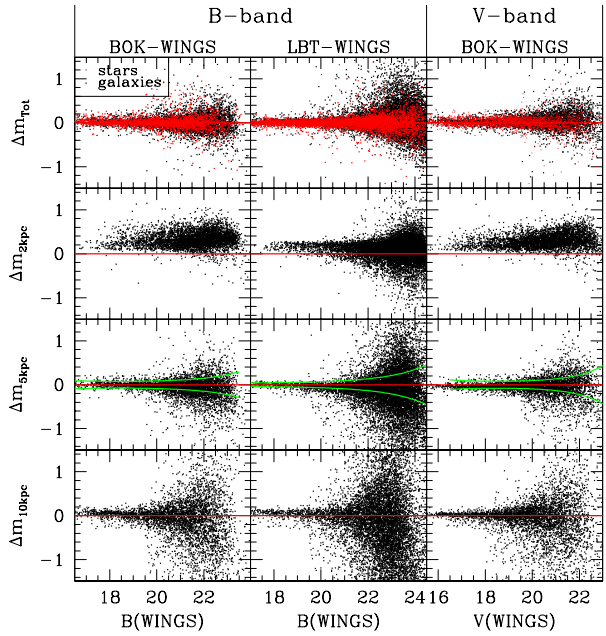}
\vspace{-3truecm}
\caption{Differences between optical bands (B,V) magnitudes from the
  catalogs presented here and the corresponding magnitudes in the
  original WINGS optical survey as a fuction of the WINGS magnitudes
  for the galaxies in common (black dots). The plots in the topmost
  row of the figure refer to the total magnitudes, while those in the
  remaining rows refer to the magnitudes within circular apertures of
  radii 2, 5 and 10~kpc, top to down. The plots in the rightmost
  column of the figure refer to the V-band, while those in the other
  columns refer to the B-band.  The red dots in the plots of the
  topmost row of the figure (total magnitudes) and the green curves in
  the plots relative to the 5~kpc apertures (third row of the figure)
  are as in Figure~\ref{internal}}
\label{external1}
\end{figure}
 
\begin{figure}
\includegraphics[scale=0.4]{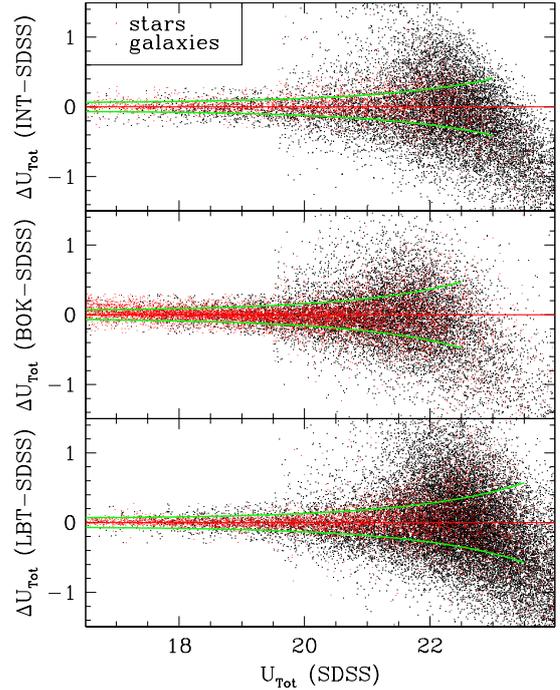}
\vspace{-0.5truecm}
\caption{Differences between U-band total magnitudes from the catalogs
  presented here and the corresponding magnitudes derived from the
  SDSS magnitudes using the the conversion formula by \cite{lupton}
  as a function of the SDSS magnitudes. Black and red dots in the
  plots refer to galaxies and stars, respectively.}
\label{external2}
\end{figure}

\section{Data quality}

\subsection{Internal comparisons}

\begin{table}
\begin{center}
\begin{tabular}{c|ccc|c}
\hline
 band & & U ($<$20.5) & & B ($<$20.5) \\
\hline
 compar.  & BOK-INT & BOK-LBT & LBT-INT  & BOK-LBT \\
 \hline   
   Total &  -0.01 ( 0.06)  &  0.00  ( 0.10)   &  0.01 ( 0.07)   &  -0.03 (0.06)\\
    2kpc & 0.00 (0.07)  &  0.09 (0.09)  & -0.09 (0.06) &  0.13 (0.07)\\
    5kpc & 0.00 (0.04 ) &  0.02 (0.06 ) &  0.00 (0.04) &  0.00 (0.07)\\
    10kpc & -0.03 (0.08) & 0.01 (0.14) & 0.00 (0.09) & -0.03 (0.08)\\  
\hline
\end{tabular}
\caption{Median values and $rms$ (in parenthesis) of the magnitude differences for internal comparisons}
\label{Tinternal}
\end{center}
\end{table}

In Figure~\ref{internal} we show the differences between magnitudes
obtained using different cameras for common objects (black and red dots
for galaxies and stars, respectively) as a function of the magnitude
of one of the two cameras, usually the one providing the most reliable
values. The green curves in the plots relative to the 5~kpc apertures
(third row of the figure) illustrate the $rms$ expected according to
the formulas reported in the headers of the catalogs (see last
sentence of Sec.~\ref{catalogs}) as a function of the magnitude. In
particular, for each couple of cameras to be compared, the $rms$
(green) curve is obtained using the error propagation rules to combine
the $rms$ formulas (reported in the catalog header) of the common
clusters for both cameras and weighting each $rms$ formula according
to the number of galaxies in the relative catalog. 
A quick report of the photometric agreement among the INT, BOK and LBT
  cameras for galaxies in common clusters is given in
  Table~\ref{Tinternal}.  Here, for each magnitude (Total, 2~kpc,
  5~kpc and 10~kpc) and for each pairwise comparison, we list the
  median value and the $rms$ (in parenthesis) of the
  magnitude differences up to a given total apparent magnitude (19.5,
  in both the B- and U-band).

From Figure~\ref{internal} and Table~\ref{Tinternal}, the
agreement among the U-band magnitudes of the different cameras turns
out to be generally good, although for the median values this
comparison test is actually meaningful just for the INT magnitudes,
since both the BOK and LBT magnitudes have been calibrated on the SDSS
magnitudes. The same holds for the BOK-LBT comparison of the B-band
magnitudes (plots in the rightmost column of the figure). In fact,
since the BOK and LBT B-band magnitudes have been both calibrated on
the WINGS optical survey, again this comparison just provides a consistency
test.

From Figure~\ref{internal}, the only remarkable (and systematic)
disagreement among the cameras concerns the magnitudes within circular
apertures of radius 2~kpc. In this case the average differences
reflect both the different average seeing conditions relative to the
different cameras and the peculiar seeing of the mosaic image of each
cluster. The influence of the seeing already disappears in the case of
the circular apertures of radius 5~kpc, for which also the scatter in
the plots turns out to be in reasonably fair agreement with the
expected $rms$ (green curves) and better than in the case of 10~kpc
aperture magnitudes.

\subsection{External comparisons}

\begin{table}
\begin{center}
\begin{tabular}{c|ccc}
\hline
  band & \multicolumn{2}{c}{B ($<$20.5)} & V ($<$19.5) \\
\hline
   compar. & BOK-WINGS & LBT-WINGS & BOK-WINGS\\
\hline    
   Total & -0.01 (0.08)  & 0.03  (0.06)   &  0.02 (0.07)\\
    2kpc & 0.27 (0.14)  & 0.16 (0.09)  & 0.21 (0.12) \\
    5kpc & 0.08 (0.07) &  0.07 (0.07) &  0.06 (0.05) \\
    10kpc & 0.01 (0.13) & 0.03 (0.14) & 0.00 (0.10) \\         
\hline\hline
 band & \multicolumn{3}{c}{U ($<$20.5)} \\
\hline
   compar. & INT-DSS & BOK-SDSS & LBT-SDSS \\
\hline
   Total & -0.05 (0.17) & 0.00 (0.20) & -0.01 (0.19) \\
\hline
\end{tabular}
\caption{Median values and $rms$ (in parenthesis) of the magnitude differences for external comparisons}
\label{Texternal}
\end{center}
\end{table}

Figure~\ref{external1} is similar to Figure~\ref{internal}, but
illustrates the quality of B- and V-band photometry through a
comparison between the magnitudes listed in the catalogs presented
here and the corresponding magnitudes from the WINGS optical
survey. It is worth recalling that, since the BOK and LBT optical
magnitudes have been both calibrated on the WINGS optical survey, this
comparison just provides an estimate of the photometric quality
(scatter) of the BOK and LBT data as a function of the WINGS
magnitudes.

For Figure~\ref{external1} one could repeat the same considerations of the
previous subsection (Figure~\ref{internal}) as far as both the
systematic disagreement of the magnitudes within circular apertures of
radius 2~kpc and the expected $rms$ as a function of the magnitude
(green curves in the 5~kpc aperture magnitudes plots).

In Figure~\ref{external2} the U-band magnitudes of the present survey
are compared with the corresponding U-band magnitudes derived from the
SDSS database using the formula provided by Lupton (2005). Again,
since the BOK and LBT U-band magnitudes have been both calibrated on
the SDSS data, the middle and bottom panels of the figure just
provide an estimate of the photometric quality (scatter) of the BOK
and LBT magnitudes as a function of the SDSS data. Instead, the
uppermost plot in the figure illustrates the fair agreement between
the SDSS and INT U-band photometric zero-points.
Table~\ref{Texternal} is similar to Table~\ref{Tinternal}, but
  refers to the external comparisons. In this case the limiting
  magnitudes adopted to compute the median values and the $rms$ of the
  magnitude differences are 19.5 for the B- and U-band, while for the
  V-band we adopt V$_{lim}$=20.5. Note that for the U-band comparisons
  we just can use the total magnitudes.

Although Table~\ref{Texternal} and Figures~\ref{external1} and
\ref{external2} refer to all clusters available for the different
comparisons, we recorded the average magnitude differences for each
individual cluster and for each magnitude (total and aperture
magnitudes). These average differences are reported in the third group
of rows of the header of each catalog (see Figure~\ref{catalog}). They
could be useful to make the catalog magnitudes (both total and
aperture magnitudes) statistically consistent with the WINGS-OPT and
SDSS magnitudes, thus allowing a more correct computation of galaxy
colors (in particular, aperture colors) for each individual
cluster. It is worth remarking, however, that significant differences
between the catalog and comparison (WINGS-OPT and SDSS) magnitudes
could persist, depending on the sizes and luminosity profiles of
individual galaxies.

\subsection{Photometric depth}

\begin{figure*}
\vspace{-2.5truecm}
\hspace{-2truecm}
\includegraphics[width=150mm,angle=-90]{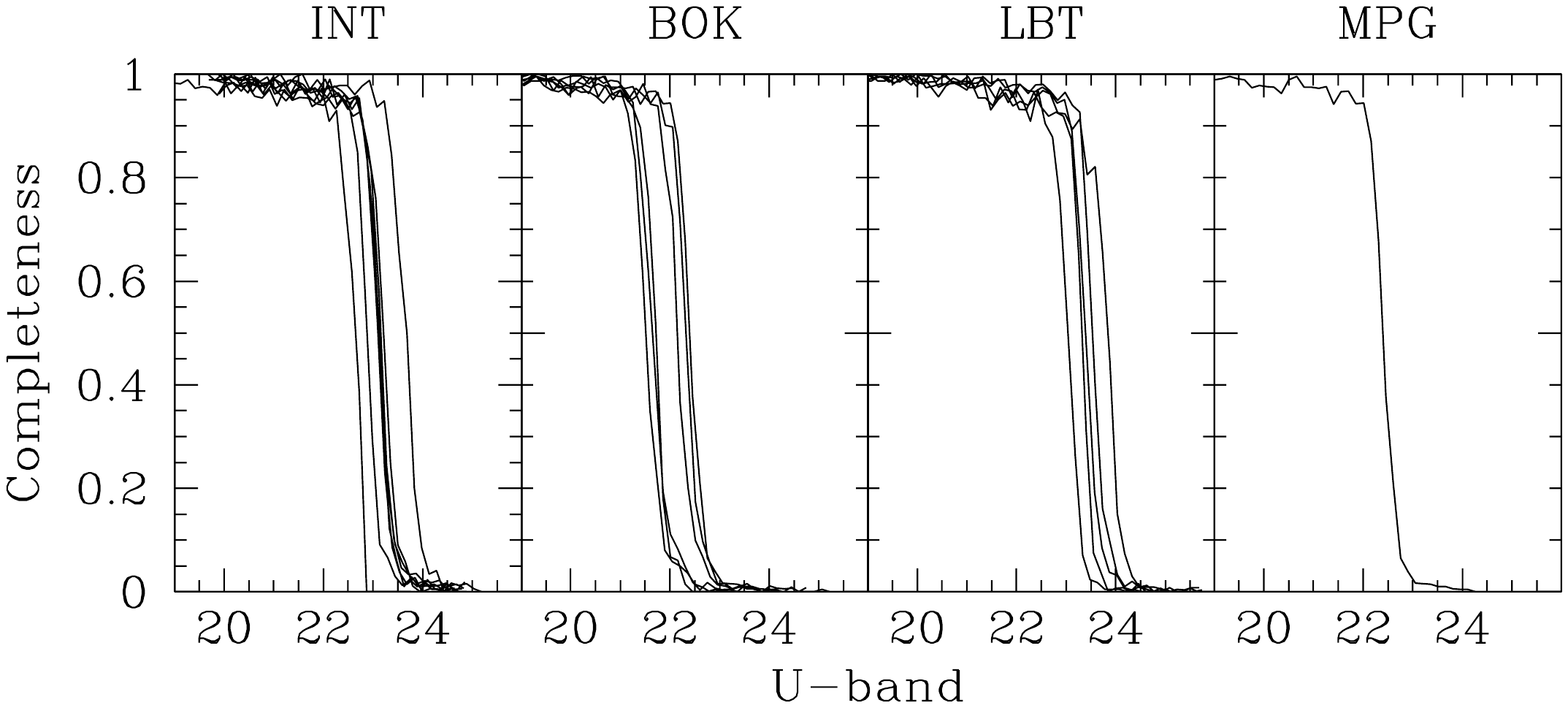}
\vspace{-4truecm}
\caption{Fractions of retrieved simulated galaxies as a function of the
  U-band total magnitudes for the four cameras.}
\label{complete}
\end{figure*}

We used simulations to estimate the detection rate in the different
bands as a function of the magnitude. In particular, for each field,
we used the IRAF task mkobj with the true image PSF to produce a
large number of artificial stars and randomly distributed upon the
field.  In particular, the artificial galaxies were simulated trying
to reproduce the magnitude, size and ellipticity distributions of the
galaxies in the real image.  Most artificial objects were simulated in
the range of faint magnitudes, where SEx is less reliable. The
artificial objects were added to the original frames and, on the new
images, SEx was run using the same configuration files used to derive
the original catalogs. The resulting catalogs were matched with the
catalogs of the artificial objects and the fraction of them recovered
by SEx in each magnitude bin was recorded. Figure~\ref{complete}
  illustrates the fractions of simulated galaxies retrieved by this
  procedure as a function of the U-band total magnitudes for the four
  cameras. Columns 4, 5 and 6 of Table~4 list, for
  each camera and filter, the range of 90\% detection rates of
  artificial galaxies. Moreover, the second group of rows in the headers
of the catalogs (see Figure~\ref{catalog}) report the 90\% detection
rates of artificial galaxies for each band, cluster and telescope. For
simulated galaxies, the 90\% detection rate in the U-band is reached
at different magnitudes for different telescopes and clusters,
spanning the intervals (22.30-23.30), (22.60-23.30) and (21.30-22.10)
for the INT, LBT and BOK imaging, respectively. The INT and LBT
telescopes provided similar results, both in terms of photometric
depth and stability. Instead, likely because of the poor weather
conditions during the two observing runs, the photometric depth of the
BOK images turns out to be worse.

\section{Conclusions}

\begin{figure}
\vspace{-0.5truecm}
\hspace{-2.5truecm}
\includegraphics[width=140mm]{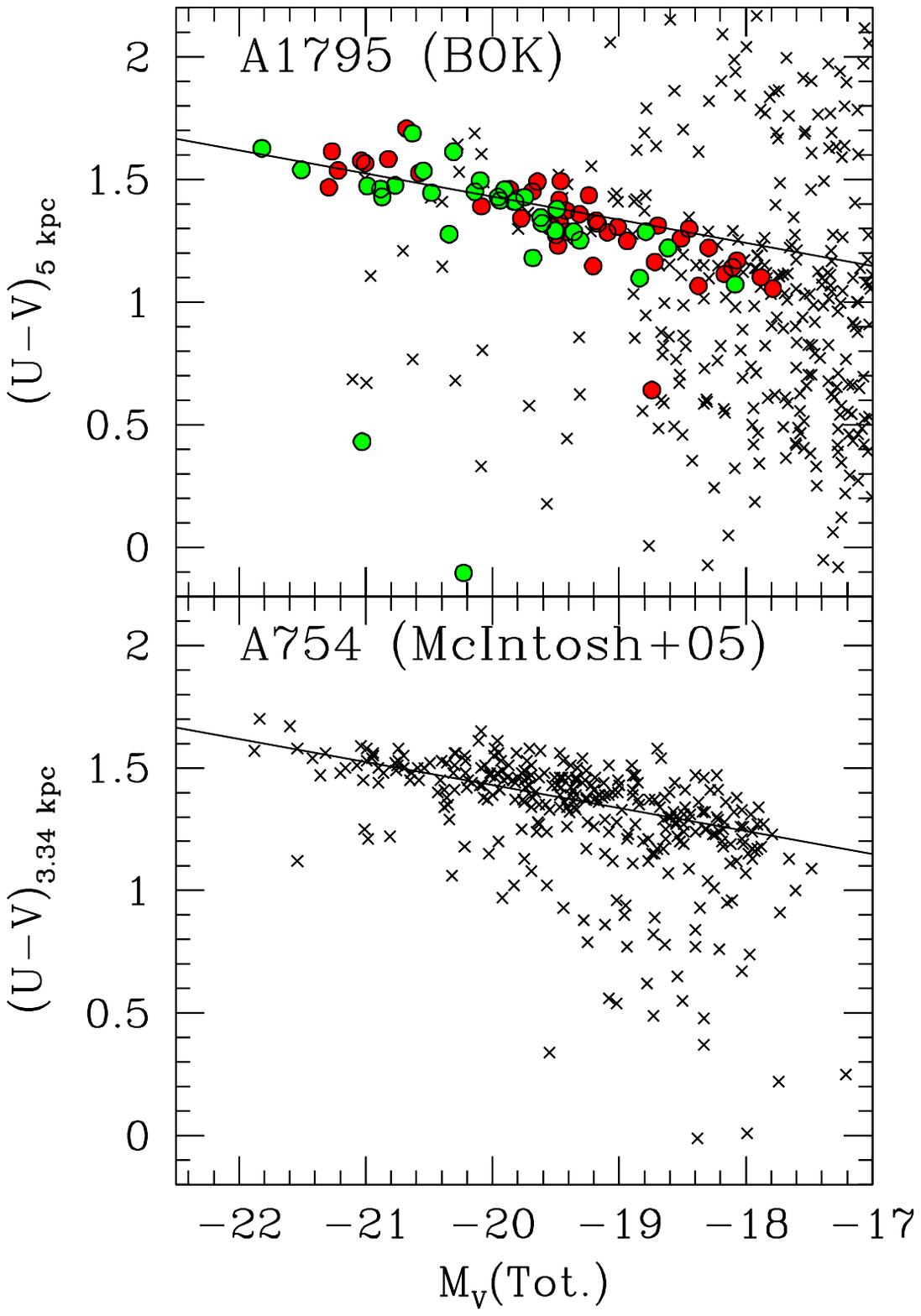}
\vspace{-4truecm}
\caption{The color-magnitude diagram (U-V)-M$_V$ for Abell~1795 (this
  work) and for Abell~754 \citep{McI}. The straight lines in the plots
  correspond to the best fit parameters given for Abell~754 in Table~8
  of \citet[][last row]{McI}. The red and green dots in the upper
  panel, respectively mark elliptical and S0 galaxies that are
  spectroscopically confirmed members of the cluster Abell~1795.}
\label{colmag}
\end{figure}

In Figure~\ref{colmag} the (U-V)-M$_V$ color-magnitude diagram (CMD) of
  galaxies in Abell~1795 (upper panel, this work) is compared with the
  same diagram obtained for Abell~754 by McIntosh et al. (2005, lower panel in the
  figure). For Abell~1795, the colors refer to the 5~kpc
  apertures and the galaxies reported in the figure are those in
common between the WINGS-OPT and BOK catalogs and morphologically
classified by Fasano et al. (2012). The U- and V-band magnitudes are
taken from the BOK and WINGS-OPT catalogs, respectively.  The red and
green dots in the plot respectively represent the elliptical and S0
galaxies spectroscopically confirmed members of the cluster.

In spite of the uncertain data quality of the BOK imaging, the red
sequence of the A1795 cluster members turns out to be very well
defined. The two very blue S0 galaxies in the plot have been visually
inspected on the images.  One of them turned out to be a close merger,
the other one being very close to a bright star.  Instead, the
elliptical galaxy below the red sequence turns out to be undisturbed,
but it is quite small (dwarf-like).

For Abell~754, the colors in the CMD refer to an aperture of
  3.34~kpc and the straight line in the plot corresponds to the best
  fit parameters given for this cluster in Table~8 of McIntosh et al. (2005, last
  row). The same straight line is also reported in the upper
  panel of Figure~\ref{colmag}. Although the CMDs are obtained with
  different apertures and for different clusters, the agreement of the two
  CMDs is quite satisfactory, thus making us confident about the good
  quality of our U-band photometry.

In a following paper of the series we will present and analyse the
color-magnitude diagrams and the average color as a function of both
the cluster-centric distance and the morphological type for all
clusters belonging to the present sample and for the clusters included
in our ongoing OmegaCam@VST survey (6 clusters already imaged in
the U-band). Moreover, both sets of data will be used to validate the
star formation rates and histories already obtained for WINGS galaxies
\citep{fritz11} and to study the color map of individual galaxies at different 
distances and position angles with respect to the cluster center. This will
help to discriminate between the various physical processes possibly 
responsible of the gas depletion in galaxies infalling into the clusters.

\begin{acknowledgement}
  We acknowledge partial financial support by contract PRIN/MIUR 2009:
  ``Dynamics and Stellar Populations of Superdense Galaxies'' (Code:
  2009L2J4MN) and by INAF/PRIN 2011: ``Galaxy Evolution with the VLT
  Survey Telescope (VST)''.
\end{acknowledgement}

\end{document}